\documentclass[twocolumn,showpacs,floatfix,pre,superscriptaddress,numbers,sort&compress]{revtex4-2}
\RequirePackage{lineno}

\usepackage{graphicx} 
\usepackage{dcolumn} 
\usepackage{amsmath}
\usepackage{mathrsfs}
\usepackage{euscript}
\usepackage{calrsfs}
\usepackage{setspace}
\usepackage{amssymb}
\usepackage{bm} 
\usepackage{physics}

\usepackage[dvipsnames]{xcolor}

\usepackage[toc,page]{appendix}
\usepackage{natbib}

\begin{document}


\title{
Collineations of particles as a tool to study nucleation and crystallization processes in liquids}

\author{V.A.~Levashov}
\affiliation{Technological Design Institute of Scientific Instrument Engineering, 630055, Novosibirsk, Russia. E-mail: valentin.a.levashov@gmail.com}


\begin{abstract}
We suggest several methods that can be used for identification of crystal nuclei in the studies of 
the nucleation and crystallization processes in multicomponent systems and in situations when 
the forming crystalline phase is different from the FCC, HCP or BCC lattices. 
The approaches are based on consideration of collineations of particles. 
They allow finding nuclei and separating them from each other.
We demonstrate the methods by considering two systems. 
For a single-component system, we perform a comparison between 
the results 
derived from the collineations and those obtained using 
the coarse‑grained bond‑orientational order parameters. 
Further, we illustrate the methods by considering the crystallization process in
a Wahnstr\"{o}m binary $BA_2$ system forming a 
rather complex crystal structure.
\end{abstract}



\today
\maketitle


\section{Introduction}\label{s:intro2}

Characterization of the structural evolution of liquids (SEL) 
on cooling at the atomic scale has 
been and remains an important research direction in computer modeling (CM). 
The details of the SEL are crucial 
for the understanding of the macroscopic behavior of liquids and for designing solid materials 
with desired properties 
\cite{kelton2010nucleation,stukowski2014computational,sosso2016crystal,
tanaka2022roles,zhang2024inhibition,papia2026measurement,
tanaka2019revealing,russo2016crystal}.
A significant branch of these investigations is related to the studies of 
crystallization and nucleation processes (CNP) 
\cite{kelton2010nucleation,karthika2016review,sosso2016crystal,tanaka2022roles,
zhang2024inhibition,papia2026measurement,russo2016crystal,blow2021seven,martirossyan2024local,gispen2025bcc}.
In particular, in CM studies of the CNP, 
it is often necessary to find those particles that 
are already in a crystalline phase and also determine the types of the formed
crystalline particles' environments 
\cite{richard1999geometrical,kumar2005voronoi,anikeenko2007polytetrahedral,
yuan2022three,gao2021fast,steinhardt1983bond,Duijneveldt1992computer,ten1995numerical,
rein1996numerical,aste2005geometrical,
lechner2008accurate,stukowski2012structure,mickel2013shortcomings,
sosso2016crystal,spellings2018machine,tanaka2019revealing,tanaka2022roles,
honeycutt1987molecular,xiao2006comparative,urrutia2016common,thuy2025influence,
kelchner1998dislocation,li2003atomeye,tsuzuki2007structural,radhi2017identification,
fang2010atomistic,malins2013identification,royall2026polymorph,wu2015hidden,
wu2023unsupervised,becker2022unsupervised,van2012parameter,gispen2025bcc,larsen2016robust,defever2019generalized,gispen2025bcc}.

There are many methods developed 
for the studies of SEL at the atomic scale 
and, in particular, methods used for the studies of the CNP. 
There are also different variations of these methods.
Still, new methods continue to emerge.  

The well known methods for the studies of SEL and CNP 
that allow addressing local structural heterogeneity
include:
Voronoi and Delaunay tesselations 
\cite{richard1999geometrical,kumar2005voronoi,anikeenko2007polytetrahedral,yuan2022three,gao2021fast},
bond-orientational order parameters (BOOPs) 
\cite{steinhardt1983bond,Duijneveldt1992computer,ten1995numerical,
rein1996numerical,richard1999geometrical,aste2005geometrical,lechner2008accurate,stukowski2012structure,mickel2013shortcomings,
sosso2016crystal,spellings2018machine,tanaka2019revealing,tanaka2022roles,gao2021fast,zeng2022bond,martirossyan2024local}, 
atomic-level stresses 
\cite{srolovitz1981local,chen1988local,levashov2008equipartition,levashov2013origin,levashov2017contribution},
common neighbor analysis 
(CNA) \cite{honeycutt1987molecular,xiao2006comparative,stukowski2012structure,urrutia2016common,larsen2020revisiting,thuy2025influence}, 
centrosymmetry parameter \cite{kelchner1998dislocation,li2003atomeye,stukowski2014computational,larsen2020revisiting},
angular distribution analysis \cite{ackland2006applications}. 

The newer methods may represent significant improvements of the older techniques or may not
be explicitly related to previous approaches.
The examples of the former include analysis of the common atomic neighborhood \cite{tsuzuki2007structural}, 
predominant common neighbor analysis (PCNA) 
\cite{radhi2017identification}, 
adaptive common neighbor analysis \cite{stukowski2012structure}, 
coarse-grained BOOPs \cite{lechner2008accurate}. 
As examples of the latter, the following can be considered: 
atomistic cluster alignment method \cite{fang2010atomistic}, 
topological cluster classification approach \cite{malins2013identification,royall2026polymorph,ingebrigtsen2019crystallization},
the approaches based on consideration of structural homology \cite{wu2015hidden,wu2023unsupervised,becker2022unsupervised}, 
polyhedral template matching \cite{larsen2016robust},
and machine learning techniques \cite{wu2015hidden,schoenholz2016structural,reinhart2017machine,defever2019generalized,alcobaca2020explainable,
becker2022unsupervised,wu2023unsupervised,takahashi2020mining,takahashi2023molecular,martirossyan2024local}.

Some of the mentioned techniques, 
when employed for the studies of the CNP,
assume that the crystal structures which can form
are relatively simple.
In the case of single-component systems,
these are often FCC, BCC, or HCP. 
Thus some of these methods, like BOOPs, are mostly suitable 
for the single-component systems that form 
relatively simple crystal structures.
Others, like CNA, can, probably, 
be tuned for addressing the developments of 
more complex structures if these target structures are 
assumed to be known.

In the case of multicomponent systems, the process 
of homogeneous crystallization often is significantly more complex 
than in the case of single component systems 
\cite{karthika2016review,sosso2016crystal,tanaka2022roles}. 
There are many phenomena that, in general, 
can cause this complexity. 
Here, we mention some of them. 
One of these phenomena is expressed 
in the eutectic phase diagram that is relevant 
to certain combinations of elements 
\cite{nishio2024liquidus,porter2009phase,pedersen2018phase,ingebrigtsen2019crystallization}. 
Crystallization in such systems is associated 
not only with structural, 
but also with chemical reorderings 
\cite{russo2016crystal,pedersen2010geometry,pedersen2018phase,
ingebrigtsen2019crystallization,nishio2024liquidus}. 
Another complexity, which may exist even in the case of congruent crystallization, 
is associated with the complexity of the intermediate phases or precursor regions 
in relation to the Ostwald's step rule which often 
plays an important role in the process of crystallization \cite{karthika2016review,truskett1998structural,o2005structure,jungblut2016pathways}. 
This behavior is related to the possible competition between different 
crystalline phases which have similar values of the free energy 
\cite{desgranges2007polymorph,royall2026polymorph} and 
also to the kinetic mechanisms of the crystallization processes \cite{sosso2016crystal,gao2021fast,jungblut2016pathways}. 
This complexity is also related to the fact 
that the variety of crystal structures that form in 
multicomponent systems greatly exceeds the variety of structures 
that usually form in single component systems 
\cite{frank1958complex,frank1959complex,pedersen2007crystallization}.

The previous paragraph indicates that methods developed 
for structural investigations of crystallization processes in single 
component systems may not be sufficient to study multicomponent systems 
and this motivates the development of methods that do not 
rely on assumptions about the final crystal structures.

Here we suggest several methods that can be used 
for the search of crystal nuclei 
with structures more complex than FCC, BCC, or HCP.
These methods are based on considerations of collineations of particles that should occur in crystalline
regions due to the (limited) translational periodicity. 
The drawback of the methods is that they cannot by themselves determine the types of the formed crystal structures.
They can only find crystalline regions in systems possessing some degree of translational periodicity. 
Thus, the determination of the types of the formed crystal structures would require an additional step.
This situation, however, is totally expectable,
as the determination of the types of the crystal structures
is a well-known problem.
In this context it should be noted that there are several programs
that can be used for this purpose
if the coordinates of particles are known. 
Previously,
for this purpose we used \cite{levashov2017crystalline} 
the FINDSYM program \cite{stokes2005findsym,findsym,isotropy}.
However, there are several other programs that also can be used.
Some of them are mentioned in Ref.~\cite{togo2024spglib}.

The idea behind the suggested methods up to a certain degree
resembles the idea behind the centrosymmetry parameter analysis 
\cite{kelchner1998dislocation,li2003atomeye,stukowski2012structure,larsen2020revisiting}.
However, the implementation of the idea is quite different.
We note that recently it has been argued that centrosymmetry
is related to the properties of ultrastable 
glasses in 2D \cite{zaccone2026ultrastable}.

Previously, we published a paper in which we investigated
the structural evolution of Kob-Andersen binary liquid
on supercooling with collineations of particles \cite{levashov2025collineations}.
There it was demonstrated that the number of collineations increases on supercooling.
Thus, it might be reasonable to suggest that the degree of translational periodicity in liquids increases
on supercooling. 
However, we essentially did not observe the clustering of the collineations.
The method of searching for collineations 
that we use here is the same as that used in Ref.~\cite{levashov2025collineations}.

The paper is organized as follows. 
In the next section, we describe the methods that we suggest for the studies
of CNP (\ref{sec:Methods}). 
Then, in sections (\ref{sec:sec-single-component}) 
and (\ref{sec:binary-system}) we describe the obtained results
for the single-component and binary systems. 
We conclude in section (\ref{sec:conclusion}).

In the supplemental materials (SM) for this paper, 
we provide some additional details concerning the
single and binary systems that we studied 
(including some dump structure files for the binary system).
Additionally, in the SMs, 
we provide several programs that can be used 
to search for crystallites starting from the
LAMMPS output (dump) structure files.
Each program is accompanied by a manual
explaining its usage. 
These manuals and programs were generated by 
the large language model DeepSeek and they
can also serve as a basis for generating 
further programs.

\begin{figure}
\begin{center}
\includegraphics[angle=0,width=3.3in]{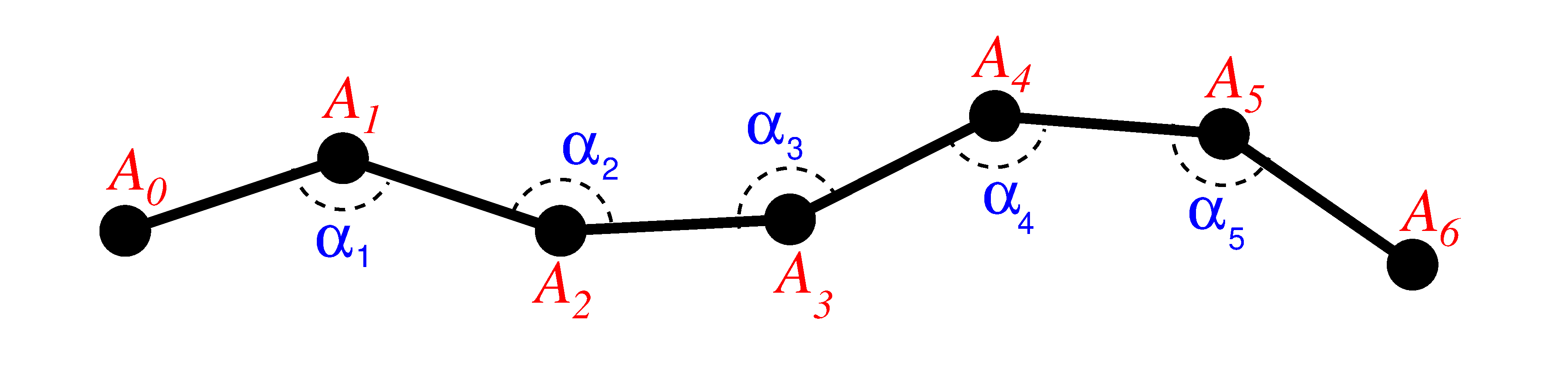}
\caption{A 2D illustration for the definition of collineations.
}\label{fig:collineation-definition}
\end{center}
\end{figure}

\section{A statement on the use of AI and Large Language Models}

We acknowledge the use of large language models (LLM) as computational 
and editorial assistants during the preparation of this work. 
In particular, 
the author used DeepSeek (DeepSeek AI), ChatGPT (OpenAI), and Yandex GPT. 
The assistance included writing and debugging of the analysis programs used 
to identify collineations and crystallites,
scientific discussions on the interpretation of the results, 
and editing of the manuscript text.
While earlier versions of some programs were developed by hand,
the final versions of the programs used in this 
study were developed with the assistance of large language models (LLMs), 
primarily DeepSeek.
In addition, the user manuals for these programs were written 
by DeepSeek and are provided in the Supplemental Materials essentially 
as received, without manual editing by the author.
The author has verified and tested the programs 
and has checked the results for consistency.
The results obtained with the final programs have also been compared with 
those obtained using earlier versions of the programs.

\section{Desciption of the developed methods \label{sec:Methods}}

The central idea of the suggested approach is to use the translational 
periodicity present in crystal structures to search for crystal nuclei in liquids. 
It is clear that because of limited translational periodicity present in nuclei, 
there will be approximately collinear arrangements of particles in them.

Below we will describe several methods for the crystalline 
nuclei search and analysis based on considerations 
of collineations of particles. 
Later, the results from these methods will be compared, 
for the single-component system,  
with the results obtained using the very common approach based 
on considerations of the coarse-grained bond-orientational order parameters (CG-BOOPs). For completeness of the descriptions of the used methods,  
at the end of this section,
we also describe the method of CG-BOOPs.

Of course, the main reason for using the methods based on considerations of collineations is the possibility to use these methods for the studies of crystallization in multicomponent 
systems with complex crystal lattices.
These methods can also be useful for the studies of single-component systems with lattices more complex than FCC, HCP, or BCC.

\subsection{Collineation Search}

Figure~\ref{fig:collineation-definition} is an illustration for the definition of collineations.
The $C(\mathscr{N}, \xi_{\text{max}})$ collineation is defined
as a configuration where $\mathscr{N}$ particles are arranged 
in such a way that each particle 
in this collineation has one or two nearest neighbors that are also part 
of the same collineation.
If a particle in the collineation is located at either end of the collineation, 
it has only one nearest neighbor that belongs to this collineation. 
If a particle in the collineation is not at the ends of the collineation, 
it has two nearest neighbors that also belong to the collineation.
For a particle in the collineation which is not at the ends of the collineation, 
the smaller angle $\alpha$ between the bonds connecting this particle to its 
two nearest neighbors within the collineation should be sufficiently close to $180^\circ$, 
according to the chosen angle cutoff. 
Specifically, the condition should be: 
$\alpha > 180^\circ - \xi_{\text{max}}$, 
where $\xi_{\text{max}}$ is the specified cutoff value. 
The smaller the value of $\xi_{max}$, the more strict is the condition 
for a chain of particles to be considered as a
collineation of type $C(\mathscr{N}, \xi_{\text{max}})$. 
Thus, the smaller the value of $\xi_{max}$, the more collinear the collineations are.
We also note that the value of the angle $\xi_{max}$ should be sufficiently small.
Otherwise, more than one particle satisfying the condition $\alpha > 180^\circ - \xi_{\text{max}}$
can be found. In this case, the particle leading to the most collinear case can be selected.
However, this situation happens only for $\xi_{\text{max}}> 25^{\circ}$. 
Here, we do not consider such angles.
We note that for the pairs of particles of different types 
in the collineations, the corresponding bond lengths can be different.
In the numerical search for collineations, 
it is convenient to adopt the following recursive algorithm.
For a pair of nearest neighbors 
(for example, $A_{0}$ and $A_{1}$ 
in Fig.~\ref{fig:collineation-definition}),
the search for the third particle ($A_{2}$) can be performed.
Thus, if there is an $A_{2}$-neighbor of the particle $A_{1}$ such 
that $\alpha_{1} > 180^{\circ} - \xi_{max}$, then this
$A_{2}$ neighbor is the third particle in the collineation. 
Further, the pair of particles $A_{1}$ and $A_{2}$ can be used to search 
for the 4th particle in the collineation. 
Thus, for the particle $A_{3}$ to be the 4th particle in the collineation, 
the angle $\alpha_{2}$ should be sufficiently close to $180^{\circ}$.
In consideration of collineations of length $\mathscr{N}$, 
this process should be repeated $\mathscr{N}-2$ times,
which is the number of angles in collineations of length $\mathscr{N}$.
All particles that belong to the collineations of a particular 
type can be labelled 
or written into the output file for further processing.
The next step in the search for the nuclei can be made
when all particles in the system that belong to 
the chosen type of collineations are found.
There are several approaches that can be used. 
We describe them further in this section.

\subsection{Joining particles into a crystallite on the basis of their neighbor connectivity. Here, we call this method the C-PNBR approach. \label{subsec:C-PNBR}}

This approach bears similarity to the early approaches \cite{Duijneveldt1992computer,ten1995numerical,
rein1996numerical} used with the BOOPs before the method of the
CG-BOOPs \cite{lechner2008accurate} was introduced. 

In this method we consider only those particles that belong to the collineations of a particular type.
By definition, it is assumed that a particle that belongs to one of such collineations belongs 
to a crystallite if it has some number of neighbors that also belong to the collineations of the same type.

It is clear that this approach may miss some particles at the surface of a nucleos. 
Also, if in the system there are several nuclei then, even though these nuclei can be 
visually distinguished in the resulting structure file, they still effectively will be 
considered as just one nucleos from the perspective of how the output structure file is written.
Thus, further processing may be required.

\subsection{Extraction of crystallites on the basis of Centro-Symmetry. We abbreviate this method as the CS-3V (or CS-2V) approach. \label{subsec:CS-3V}}

This approach up to a certain degree resembles the method based on the consideration of the centro-symmetry parameter. In reality these two approaches are quite different. The centro-symmetry parameter is, by definition, a scalar. The method that we describe here is based on consideration of vectors.

In this approach we consider only those particles that belong to the collineations containing at least three particles. In fact, even in the liquid system, 
there are many particles through which pass several such collineations.
Thus, we can associate with each such particle several vectors whose directions are defined by the directions of such three-particle collineations. In the implementation of the algorithm it is assumed that 
vectors $+\vec{v}$ and $-\vec{v}$ are the same.

Further, we say that two neighboring particles belong to the same 
crystallite cluster if, for example, three vectors associated with one particle have directions which are sufficiently similar to the directions of three vectors associated with another particle. Performing the iterative search over the neighbors of particles that already belong to a crystallite cluster, the program will continue building a crystallite. This search over the particles belonging to a crystallite stops at some moment because of the absence of neighbors that can be added. In this case, the program starts to search for a particle that does not belong to a crystallite and that can start a new crystallite cluster. The described process can continue until all particles in the system are exhausted. 
Only clusters with sufficiently large numbers of particles 
can be written to the output structure file.

In our view, 
this approach performs quite well.
The important advantage of this approach is that it naturally distinguishes between 
different crystallite clusters. This is convenient for further processing.

One arguable shortcoming of this approach, in comparison to the previously described one and the next approach, is that it does not contain naturally in itself the characteristic lengthscale of the found clusters, 
even though the sizes of the crystallite clusters can be considered as such a characteristic.

Another arguable shortcoming of this approach is that the clusters of particles that 
it finds do not look, under visual inspection, as crystallite clusters.
It is our assumption that the approach, besides real crystallite clusters, can also find regions near such crystallites which are in pre-crystallite states.

\subsection{Method for crystallites extraction on the basis of the global analysis of the collineations' directions. 
Here, we call this method the GACD approach.\label{method:GACD}}

This approach consists of three steps. 

On the first step, after a collineation of a particular type is found,
its direction is compared to the directions associated with the previously found crystallite clusters. 
On this first step, 
each crystallite is defined through only one direction. 
If the direction of the newly found collineation is similar 
enough to one of the directions of the previously found crystallites,
then the particles that form the newly found collineation 
are classified as belonging to the crystallite.
If there is no crystallite with a direction similar enough 
to the direction of the newly found collineation, then this collineation founds a new crystallite.

On the second step, 
several other directional vectors that can be associated with 
the found crystallites are determined. 
It is clear that with any real three-dimensional 
crystallite at least three translational directions can be associated. 
However, note that in the FCC lattice there are six collineations with different directions that pass through each particle 
(of course, any three of these directions can be represented 
as linear combinations of the other three directions).
Thus, on the second step, 
we search the crystallites found in 
the first step for other possible 
directions of 
the collineations that can be formed 
by particles belonging to the crystallites.
Note that in some cases it might be necessary to consider not 
only the nearest neighbors, but also the second neighbors. 
Thus, the number of directions associated with a crystallite 
can be larger than six.
We did not try to reduce the number of found directions
to three and used a list of found directions that could contain more than
six vectors.

The third step in this approach is the merging of different 
found crystallites into one crystallite if both crystallites have some
number of directions that are sufficiently similar.

For the complex case of the binary system that we considered, 
this algorithm does not allow distinguishing different crystallites
perfectly. However, it performs reasonably well and, 
since it starts from considerations of particular collineations,
it naturally introduces the crystallites' minimum length scale.

\subsection{Extraction of crystallites using collineations as seeds.
Here, we call this approach C-Seed method.\label{subsec:C-Seed}}

In our view, it is one of the most successful approaches.

The algorithm is the following.
After a collineation of sufficient length is found 
it is used as an original seed for a crystallite.
Further, the neighbors of the particles which are 
in the original collineation are checked if they are members 
of other collineations which are 
sufficiently parallel to the original collineation.
If there are such collineations then the particles that form such collineations are added to the crystallite 
founded by the original collineation.
In this way, through consideration of the neighbors of 
the particles which are already marked as belonging to the crystallite, 
the crystallite grows until no further collineations 
sufficiently parallel to the original collineation are found.

It is worth noting that 
there are two natural ways to implement the algorithm. 
In one way, 
the directions of the neighbor-collineations are compared 
to the direction of the original collineation. 
That is how the algorithm is implemented now. 
In another approach, 
the direction of the newly found collineation, 
that starts on a neighbor 
of the particle belonging to the crystallite, 
can be compared with the direction of the collineation 
that passes through the neighboring particle that belongs to the crystallite.

\subsection{The method for merging of crystallites on the basis of their global neighbor connectivity. Here, we call this method Cryst-G-NBR-M.\label{subsec:Cryst-G-NBR-M}}

It is also possible to merge crystallites on the basis of their global neighbor connectivity. Thus, if a sufficiently large number of particles of one crystallite are the neighbors of particles forming another crystallite, then these two crystallites can be merged into one 
crystallite.

In general, the criterion for the number of neighbors that are required for crystallites merging should depend on the sizes of the crystallites.
Thus, it might be necessary to use the program iteratively with different criteria for merging.
However, this method, although it seems to be very natural to use, does not always perform well. At least we found that it is so for the binary system that we considered.

Also, this method can merge into one two neighboring crystallites with different
orientations.

\subsection{Analyze crystallites diagnostic tool}

After crystallites are determined with some method, 
it might be reasonable
to analyze their sizes and the directions of 
collineations of particular types that can be found within crystallites.
One may also want to consider the numbers of collineations associated
with a particular direction of a crystallite. 
The angles between the crystallite's directions also can be considered.
Besides providing general information about the structure of crystallites, such analysis can indicate, for example, that a particular crystallite, in fact, can contain within itself two different crystallites.

\subsection{The method of coarse-grained bond-orientational order parameters (CG-BOOPs). 
A reference method for comparison with the suggested approaches based on considerations of collineations \label{subsec:BOOPs}}

The method of the bond-orientational order parameters (BOOPs) 
is very well known in the studies of the liquid structures and 
in the investigations of the nucleation process
\cite{
steinhardt1983bond,
Duijneveldt1992computer,
ten1995numerical,
rein1996numerical,
richard1999geometrical,
aste2005geometrical,
desgranges2007polymorph,
lechner2008accurate,
stukowski2012structure,
mickel2013shortcomings,
sosso2016crystal,
spellings2018machine,
levashov2018green,
levashov2019anomalous,
haeberle2019distinguishing,
tanaka2019revealing,
tanaka2022roles,
gao2021fast,
zeng2022bond,
Usman2026208515}.
A variant of this method, 
based on considerations of the coarse-grained BOOPs (CG-BOOPs) 
introduced by Lechner and Dellago \cite{lechner2008accurate}, 
is particularly effective in the search of nucleation 
events \cite{lechner2008accurate,stukowski2012structure,sosso2016crystal,tanaka2019revealing}.
The probability distributions (PDs) of the CG-BOOPs are significantly 
narrower than the PDs of the BOOPs which are not coarse-grained.
The narrowness of the PDs of the CG-BOOPs allows for a much more 
clear distinguishing between FCC, HCP, and BCC lattices.
For this reason, we considered only the CG-BOOPs.
Here we provide definitions which are 
immediately relevant to our work.

In a particular coordinate frame, the orientation 
of the bond between a particle \( i \) and its nearest neighbor \( j \)
is represented by the spherical angles \( (\theta_{ij}, \phi_{ij}) \).
Following Steinhardt et al., the local complex bond-orientational order parameter (BOOP)
components are defined as \cite{steinhardt1983bond}:
\begin{eqnarray}
    q_{lm}(i) = \frac{1}{N_b(i)} \sum_{j=1}^{N_b(i)} 
    Y_{lm}(\theta_{ij}, \phi_{ij}),
    \quad m = -l, \ldots, l,
\end{eqnarray}
where \( Y_{lm} \) are the spherical harmonics and \( N_b(i) \) 
is the number of nearest neighbors of particle \( i \).

Then, following Lechner and Dellago \cite{lechner2008accurate},
the coarse-grained complex bond-orientational order parameter (CG-BOOP) components are defined as
\begin{eqnarray}
    \bar{q}_{lm}(i) = 
    \frac{1}{N_b(i) + 1}
    \sum_{k=0}^{N_b(i)} q_{lm}(k),
\end{eqnarray}
where the sum runs over the particle \( i \) and its nearest neighbors.

Then, the coordinate system independent (rotationally invariant) 
coarse-grained (local) bond-orientational order parameter is:
\begin{eqnarray}
    \overline{Q}_l(i) = 
    \sqrt{\frac{4\pi}{2l + 1} 
    \sum_{m=-l}^{l} \abs{\bar{q}_{lm}(i)}^2 }.
\label{eq:Ql}    
\end{eqnarray}

Another higher-order coarse-grained invariant BOOP is:
\begin{eqnarray}
    \overline{W}_l(i) & \\
    = \sum_{\substack{m_1, m_2, m_3 \\ m_1 + m_2 + m_3 = 0}}
    \begin{pmatrix}
        l & l & l \\
        m_1 & m_2 & m_3
    \end{pmatrix}&
    \bar{q}_{lm_1}(i)\, \bar{q}_{lm_2}(i)\, \bar{q}_{lm_3}(i),\nonumber
\label{eq:Wl}    
\end{eqnarray}
where \( \begin{pmatrix} l & l & l \\ m_1 & m_2 & m_3 \end{pmatrix} \)
are Wigner 3j symbols.

The normalized form that we used is:
\begin{eqnarray}
    \hat{\overline{W}}_l(i) = 
    \frac{\overline{W}_l(i)}
    {\left( \sum_{m=-l}^{l} \abs{\bar{q}_{lm}(i)}^2 \right)^{3/2}}.
\end{eqnarray}
All invariant BOOPs should vanish for completely isotropic environments.
The first non-zero values of the BOOPs occur for $l=4$ ($Q_4$ and $W_4$) for local environments with cubic symmetry.
For local environments with icosahedral symmetry, the first non-zero values of the BOOPs occur for $l=6$ ($Q_6$ and $W_6$).
The value of the parameter $Q_{6}$ is rather 
sensitive to the development of crystalline or icosahedral 
local ordering \cite{steinhardt1983bond,Duijneveldt1992computer}.

As described in the next section and SMs, 
we calculated the probability distributions of the CG-BOOPs for 
the crystal states of our single-component system at several temperatures and at a target pressure of 1 atm.
On the basis of these PDs curves (for the crystal states) and also on the basis of the 
previous considerations of the BOOPs \cite{steinhardt1983bond,Duijneveldt1992computer,ten1995numerical,rein1996numerical,desgranges2007polymorph,
lechner2008accurate,stukowski2012structure,mickel2013shortcomings,sosso2016crystal,spellings2018machine,levashov2018green,levashov2019anomalous,
haeberle2019distinguishing,tanaka2019revealing,zeng2022bond,Usman2026208515}, we adopted the following conditions 
for defining a particle to be in the BCC environment:
\begin{eqnarray}
&+0.02 < \overline{Q_{4}} < +0.06, \;\;\;\;\;
+0.0 < \overline{W_{4}} < +0.16, \;\;\;\;\; \nonumber \\
&+0.40 < \overline{Q_{6}} < +0.52, \;\;\;\;\;
+0.0115 < \overline{W_{6}} < +0.0145,\;\;\;\;\;
\label{eq:boops-conditions}
\end{eqnarray}
where $\overline{Q_{4}}$, $\overline{W_{4}}$, $\overline{Q_{6}}$, and $\overline{W_{6}}$ are the coarse-grained BOOPs defined in (\ref{eq:Ql},\ref{eq:Wl}).
In our considerations, 
if a particular particle in the system had the values of 
the CG-BOOPs in the ranges shown in (\ref{eq:boops-conditions}),
then it was considered to be a crystalline particle in the BCC environment. 
We did not apply an additional condition that for a particle to be considered 
crystalline it also should have some number of crystalline neighbors. 
This condition is usually applied when non-CG-BOOPs are considered \cite{ten1995numerical,rein1996numerical}.
We also performed analysis with this additional requirement for the CG-BOOPs.
In our view, this additional condition does not lead to a noticeable change in the results. 
Thus, here we present only the results obtained without the requirement for some number of crystalline neighbors.

\section{Single component system \label{sec:sec-single-component}}

\subsection{Description of the single component system}

We used a model of particles interacting through 
the modified Johnson pair potential (MJPP) for iron 
to demonstrate how the methods based on consideration 
of collineations compare with the CG-BOOPs method.

The pair potential introduced by R.A. Johnson was originally 
used to study the behavior of interstitials 
and vacancies in $\alpha$-iron \cite{johnson1964interstitials}.
Consequently, a modification of this potential was used, in particular,
to study the structure of melted iron and as a general model system
for the studies of the glass transition 
\cite{srolovitz1981structural,chen1988local,tomida1995molecular,
levashov2008atomic,levashov2013origin,levashov2014dependence}.

\begin{figure*}
\begin{center}
\includegraphics[angle=0,width=6.5in]{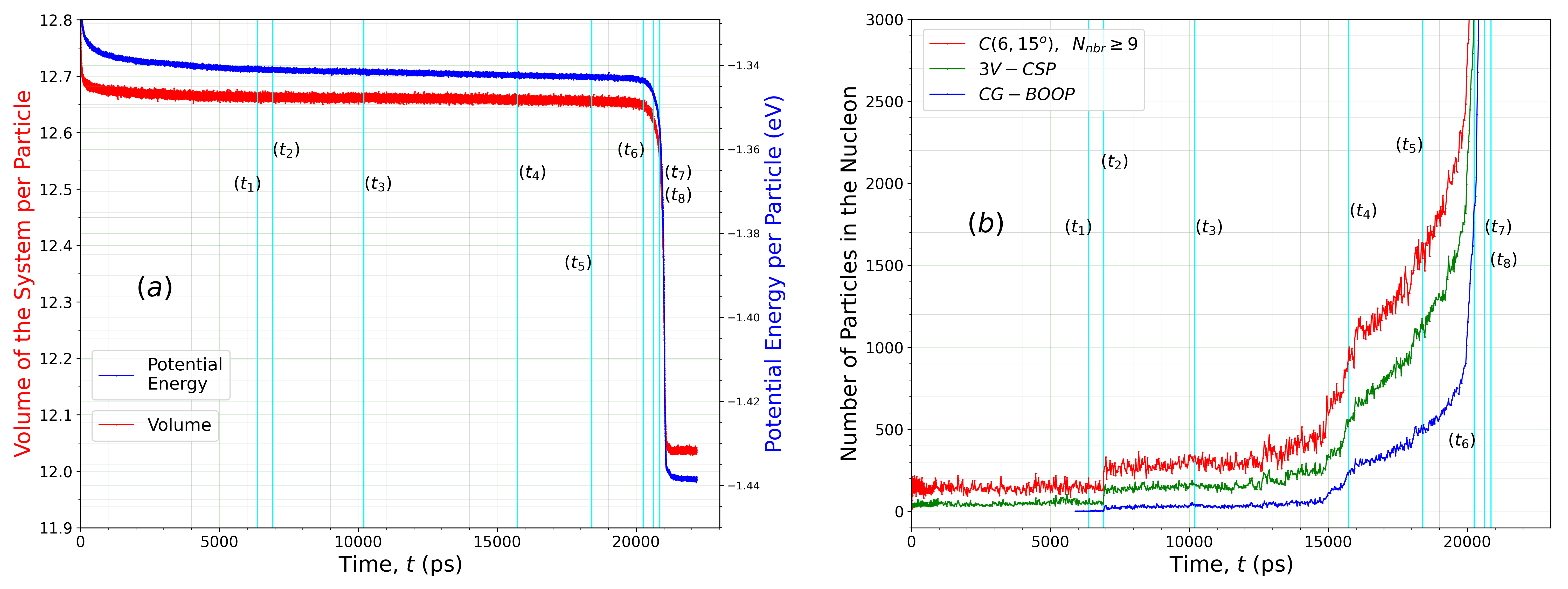}
\caption{(a) The dependencies of the volume of the system and its potential energy on time at $T=700\,(K)$ after an instantaneous 
drop in temperature from $T=1300\,(K)$. The abrupt drop in the volume of the system and its potential energy near
$t=21'000\,(ps)$ is related to the crystallization of the system. 
Figure 5 in SM shows the mean square displacement of particles as a function of time in this simulation.
At the times corresponding to the vertical cyan lines, the particles' configurations have been analyzed
with the $C(6,15^{o})$ collineations and the CG-BOOPs. The results are presented in 
Fig.~\ref{fig:Frames-LLL-01}. It follows from these snapshots that
the nucleos forms long before the abrupt drop in the volume and potential energy, i.e., at $t \approx 6'925\,(ps)$.
(b) Dependencies on time of the numbers of particles in the nucleos found by three methods: the method of $C(6,15^{o})$ collineations with at least nine neighbors for each particle in the nucleos (C-PNBR), the CS-3V method based on considerations of three-particle collineations with similar directions on the neighboring particles, and the method of CG-BOOPs. The data in this panel are for the same simulation from which panel (a) was produced. It is clear from Fig.~\ref{fig:Frames-LLL-01} that the analysis based on considerations of CG-BOOPs finds only those particles that truly belong to the nucleos, while the analysis based on consideration of $C(6,15^{o})$ collineations also finds some particles which are not part of the nucleos. However, from the comparison of the results from the two methods, it is also clear that the CG-BOOPs method does not find some particles which are related (close) to the nucleos. The results for the $CS-3V$ method (see Fig.~\ref{fig:nucleos-CSP-mJp}) suggest that it captures some pre-critical crystal-like fluctuations in the system. The curve for the $CS-3V$ method shows the size of the largest cluster only. The different numbers of particles found by the three methods are reflected in the different behaviors of the curves in (b).
}\label{fig:V-PE-Growth-01}
\end{center}
\end{figure*}

\begin{figure*}
\begin{center}
\includegraphics[angle=0,width=6.5in]{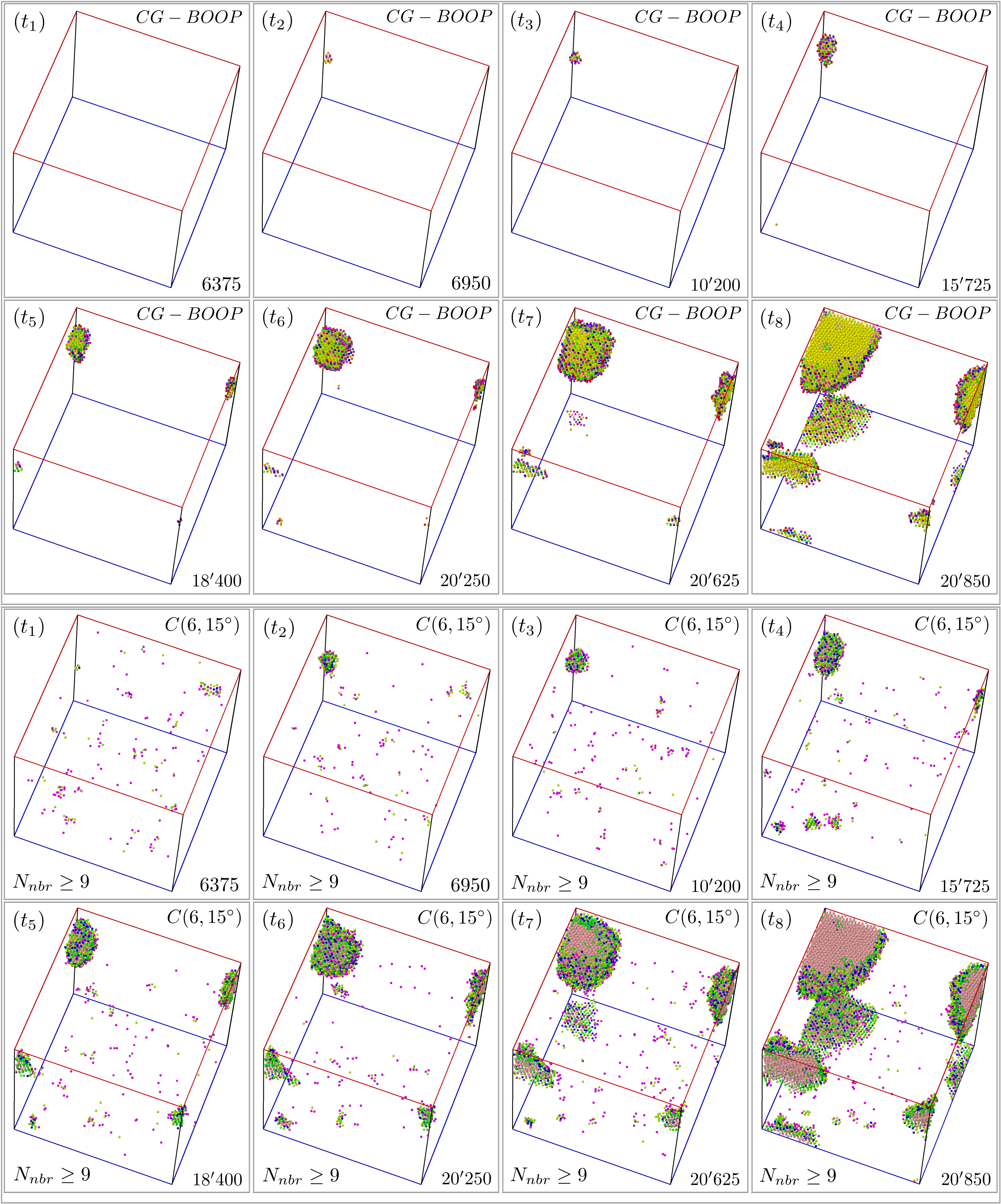}
\caption{Results for the particles interaction through the modified Johnson pair potential for iron.
The first and second rows show the result obtained using the method of coarse-grained bond-orientational order parameters 
(CG-BOOPs) at times shown in 
Fig.~\ref{fig:V-PE-Growth-01} (\ref{subsec:BOOPs}).
The time for each snapshot is indicated in the top-left corner of each panel, with its numerical value in picoseconds repeated in the bottom-right corner. The CG-BOOPs method clearly distinguishes those particles that belong to the nucleos from those that do not. Thus, additional analysis of the particles' environment is not required. However, in the figure, we colored particles according to the number of their neighbors that also belong to the nucleos.
The third and forth rows show the results from 
the C-PNBR method (\ref{subsec:C-PNBR}).
Snapshots of the configurations processed with the $C(6,15^{\circ})$
collineations are shown. Only those particles are shown that have at least nine neighbors that also belong to the $C(6,15^{\circ})$ collineations. 
The times corresponding to the panels are the same as those 
in the first and second rows. 
Thus, this results can be directly compared with the results in the first and second rows. The size of the nucleos found with the C-PNBR method is clearly larger than the size of the nucleos found with the CG-BOOPs method. 
However, this method also finds some particles that do not belong to the major nucleos. 
It is possible that these 
particles can be associated with the pre-critical fluctuations. 
}\label{fig:mJp-nucleos-BOOPs-and-lll-collin}
\end{center}
\end{figure*}

\begin{figure*}
\begin{center}
\includegraphics[angle=0,width=6.5in]{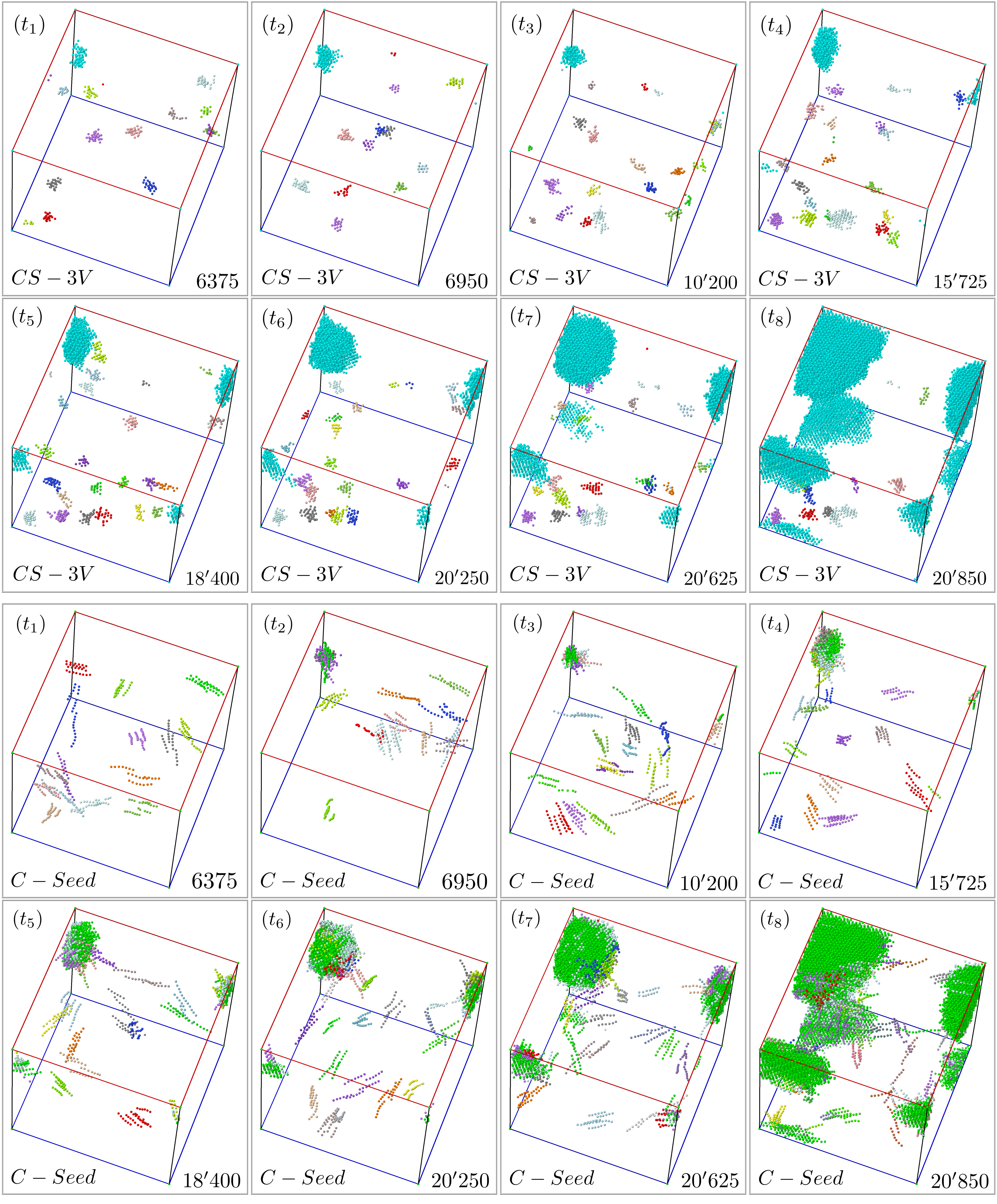}
\caption{In the first two rows the results from the CS-3V method are shown (\ref{subsec:CS-3V}).   It is of interest that this method also finds additional small regions that might represent the pre-critical fluctuations. 
This figure can be directly compared with 
Fig.~\ref{fig:mJp-nucleos-BOOPs-and-lll-collin}.
In Fig.~6 of the SM for the single component system the results obtained with the
CS-2V approach are also shown. 
Thus, the results obtained with the CS-3V and CS-2V approaches can be compared.
(b) Results obtained with the C-Seed method (\ref{subsec:C-Seed}). 
The collineations used as seeds had to
be a $C(6,15^{\circ})$ collineation. 
Then, for a collineation to join a crystallite it also had to be a
$C(6,15^{\circ})$ collineation.
If the directions of two collineations were within $15^{\circ}$ then their directions were considered as similar enough. Only clusters containing at least 25 particles are shown.
Note that in the time interval between
$(t_2)$ and $(t_3)$ the nucleus essentially does not grow. Then, in the interval between $(t_3)$ and $(t_6)$ the nucleus grows rather slowly, considering the length of this time interval.
}\label{fig:nucleos-CSP-mJp}
\end{center}
\end{figure*}

The details concerning the MJPP are provided in the SMs. Here we only state some of its properties that are necessary for understanding the results. The potential has a minimum at $r_{min} \approx 2.617\;\AA$ 
with depth $U_{min} \approx -0.252\;(eV)$ or $2924\;(K)$.
The potential has continuous first and second derivatives and goes to zero at $r_{o} = 3.44\;\AA$. 
As described in the SMs, this potential, for all positive pressures at $T=0\;(K)$, 
favors the BCC crystal lattice over the FCC lattice.
For these parameters of the potential and the iron atoms, the characteristic time is:
$\tau \equiv (m \sigma^2/\epsilon)^{1/2} \approx 0.126$ (ps).
The characteristic pressure is: $\pi \equiv (\epsilon/\sigma^3) \approx 225$ (GPa).

We performed canonical NPT simulations using the LAMMPS molecular dynamics (MD) program \cite{Plimpton1995,thompson2022lammps,lammps}. 
The temperature and pressure were introduced using the Nos{\'e}-Hoover thermostat and barostat \cite{nose1984unified}.

The particles were placed in a cubic simulation box with periodic boundary conditions (PBC) in all three directions. 
The number of particles in the box was  $N=128'000$.
The simulations were performed with the target pressure corresponding to 1 atm. Thus, our simulations were performed essentially at zero pressure from the perspective of the magnitude of interatomic forces.

In our MD simulations we observed the typical behavior when the crystal can be significantly overheated and the melted state can be significantly supercooled (SM). We did not search for the precise value of the melting temperature. More details concerning the system preparation and its behavior are provided in the SMs. 
Here we mention only that for the crystal states existing in the temperature interval, $100\;(K)\; \leq T\; \leq 2000\;(K)$, the first minimum in the pair distribution function (PDF) is close to $r_c = 3.5\;\AA$ (Fig.~3(f) of the SMs), 
while
in the liquid's (glass's) state the first minimum 
in the PDF occurs at $r_{cutoff} = 3.34\;(\AA)$. 
For the crystal state at low temperature ($T=700$ (K)),
the nearly flat region starts
at this distance. 
We used the value of $r_{cutoff} = 3.34\;\AA$ to define the (1st) nearest neighbors.
In the search for the nuclei, 
this cutoff distance was used for all methods, i.e., for the methods based on considerations of collineations 
and for the method based on consideration of the CG-BOOPs.

With the value $r_c = 3.34\;\AA,$ we calculated the probability distributions (PDs) for the selected CG-BOOPs at several 
temperatures for the crystal state (Fig. 4 (SM)). 
From these PDs, the conditions for the particles to be in the BCC state were adopted. 
We already presented these conditions in section (\ref{subsec:BOOPs}).

We observed crystallization of the liquid state in the temperature interval $700\;(K)\; \leq T\; \leq 1250\;(K)$ (SMs). 
In agreement with the general expectation from the time-temperature-transformation (TTT) curve (C-curve), 
the formation time of the critical nucleus at $T = 1250\;(K)$ and $T = 700\;(K)$ 
can be rather large \cite{kelton2010nucleation,karthika2016review}.

Here, to compare the results obtained by the four approaches for the nucleos search, 
we discuss the nucleation process at $T = 700\;(K)$. 
The starting configuration for the simulations at $T = 700\;(K)$ 
was a particular liquid equilibrium configuration at $T = 1300\;(K)$. 
Starting from this configuration, we instantaneously dropped 
the target temperature in simulations to $T = 700\;(K)$.

The lengths of the simulation box sides at $T=700$ (K) 
were $L \approx 117.47\,(\AA)$ 
before crystallization and $L \approx 115.49\,(\AA)$ when 
nearly all of the system crystallized.

\subsection{Results for the single-component system}

In panel (a) of Fig.~\ref{fig:V-PE-Growth-01} the dependencies on time 
of the volume of the system per particle and the potential energy of the system per particle are shown.
In the figure, the initial fast relaxation ends 
at $t \approx 1000 (ps)$ and it is followed by 
a much slower relaxation 
that can also be associated with 
the ageing process in glasses.
We note that for $T=700\,(K)$ 
the waiting time 
for the formation of a critical nucleos 
is rather large. 
Thus, it is not surprising that 
we observe the formation of only one nucleos.
Of course, for the purpose of demonstration, we intentionally chose the temperature that 
would allow us to observe the nucleation process clearly.
The fast crystallization process in the system 
starts at $t \approx 21'000 (ps)$.

During the relaxation process we monitored the structures formed by the particles found 
by the four methods.
In particular, we used 
the 
BOOPs (\ref{subsec:BOOPs}), 
C-PNBR (\ref{subsec:C-PNBR})
CS-3V (\ref{subsec:CS-3V}), and the 
C-Seed (\ref{subsec:C-Seed})
methods.

The vertical (cyan) lines 
in Fig.~\ref{fig:V-PE-Growth-01}(a,b), 
marked with $(t_1,...,t_{8})$,
show the selected times at which the 
particles forming nucleos
are shown in 
Fig.~\ref{fig:mJp-nucleos-BOOPs-and-lll-collin},\ref{fig:nucleos-CSP-mJp}.
The increases with time in the numbers of particles 
belonging to the nucleos, 
as found by 
the 
C-PNBR (\ref{subsec:C-PNBR}), 
CS-3V (\ref{subsec:CS-3V}), and
BOOPs (\ref{subsec:BOOPs}) 
methods are shown 
in Fig.~\ref{fig:V-PE-Growth-01}(b). 

In our search for the nuclei,
within C-PNBR (\ref{subsec:C-PNBR}) approach, 
we tried collineations of different lengths. 
We also tried different numbers of the collineation-neighbors.
We found that the following conditions for a particle to be in the nucleos work quite well: 1) the particle belongs to a collineation of type $C(6,\,15^{o})$.
2) the particle has at least 9 nearest neighbors that also belong to collineations of type $C(6,\,15^{o})$.
The results for this choice of conditions are presented in 
Fig.\,\ref{fig:mJp-nucleos-BOOPs-and-lll-collin}. 
This approch in its spirit is somewhat similar to the approach
based on consideration of the BOOPs \cite{Duijneveldt1992computer,ten1995numerical,rein1996numerical}.

The major result that becomes clear from the comparison of corresponding panels in
Fig.~\ref{fig:mJp-nucleos-BOOPs-and-lll-collin}
is that the nucleos found by the method of collineation
of particles is larger than the nucleos found by the method of CG-BOOPs.
This result is also supported by the curves in Fig.~\ref{fig:V-PE-Growth-01}(b).
Thus, the method of collineation of particles can be useful in identifying the regions
of the system with a modified or more crystal-like structures. 
It should also be noted that the method of collineations finds some particles 
which do not belong to the nucleos. 
It is possible that the method of collineations captures structural fluctuations in the liquid associated with the precritical nuclei, whereas the method of CG-BOOPs does not capture these fluctuations.

Another noticeable observation from Fig.~\ref{fig:V-PE-Growth-01}(a,b) and Fig.~\ref{fig:mJp-nucleos-BOOPs-and-lll-collin}
is that the initial nucleos forms at $t \approx 6375$ (ps). 
Then, in the interval of time $t \in (6950\,:\,10'200)$ (ps) the nucleos essentially does not grow.
Further, in the interval of time $t \in (10'200\,:\,20'250)$ (ps) the nucleos grows rather slowly, 
in comparison to the very fast crystallization 
that starts at $t \approx 20'625$ (ps).
Thus, our results show that in some ranges of the nucleos size (when it already clearly exists)
the nucleos grows very slowly (or nearly does not grow). 
We observed this behavior in the very supercooled liquid or in the glass state. 
We did not observe this behavior in the ``high-temperature" supercooled liquid, 
for example, at $T=1250$ (K). 

\subsection{Conclusion for the section on single-component system}

In this section, we demonstrated for the simple system the possibility of investigating the nucleation processes with the methods based on consideration of collineations of particles.
The considerations of this simple system
allowed us to make a valuable comparison with the results
obtained with the method of CG-BOOPs (\ref{subsec:BOOPs}).
 
Of course, 
the major interest in the methods based on consideration of collineations of particles is in the possibility to address the behaviors of systems forming the crystal lattices which are more 
complex than the FCC, BCC, or HCP lattices. 
This is especially relevant in considerations of the 
multicomponent system.
In the next sections, we discuss a particular example of such a system. 

Because our goal is to demonstrate the method, we do not discuss here the results of our observations in the context of the nucleation theories. 
However, 
we note that we observed an interesting, 
though expected, 
behavior for the nuclei formation and growth 
at temperatures $T=750$ (K) and $T=700$ (K). 
In particular, at these temperatures nuclei 
containing up to 100 or 200 particles 
can exist in the system for a long time essentially without growing.

\section{Binary system \label{sec:binary-system}}

\subsection{Description of the binary system}

\begin{figure*}
\begin{center}
\includegraphics[angle=0,width=6.5in]{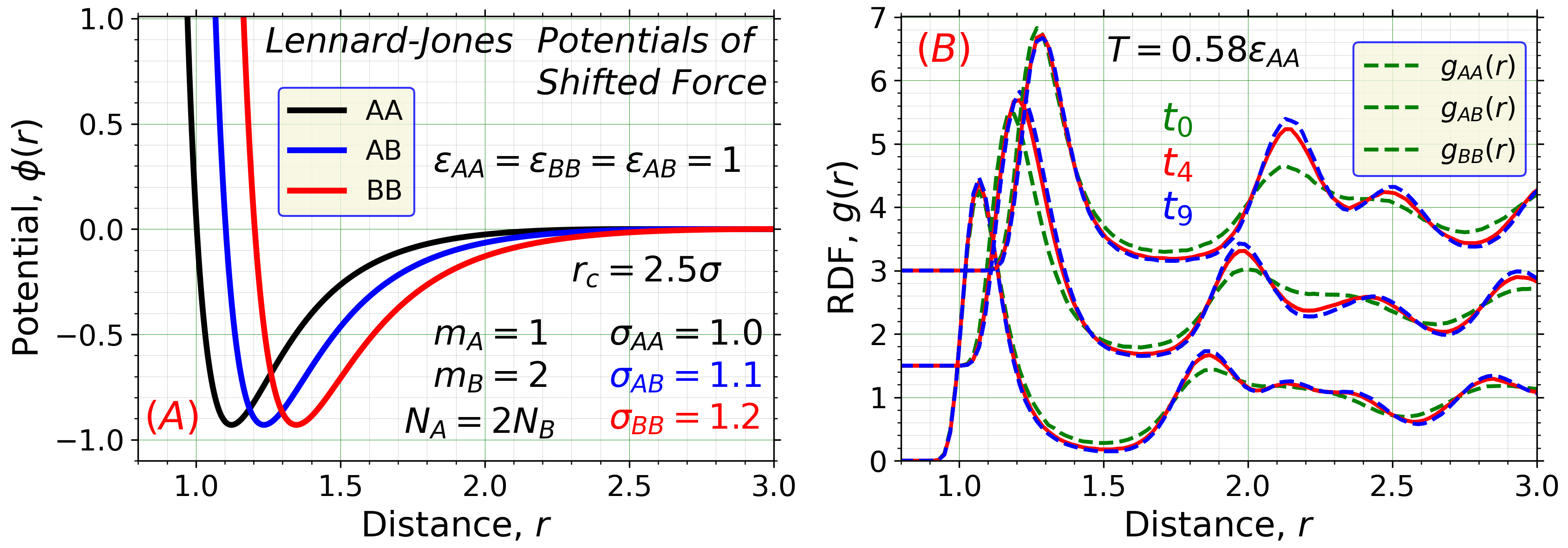}
\caption{
(A) The interaction potentials and their parameters 
for the binary particle system under consideration.
(B) Partial pair distribution functions (PPDFs) for three different time intervals in the process of crystallization.
The lowest three curves correspond to $AA$ (or $ZnZn$) pairs. The curves shifted upward by 1.5 and 3.0 show results for
$AB$ (or $MgZn$) and $BB$ (or $MgMg$) pairs.
The green dashed, red solid, and blue dashed curves were obtained 
from the structures at times $t_{0}$, $t_{4}$ and $t_{9}$ which
correspond to the time intervals $(0)$, $(4)$, and $(9)$ in Fig.~\ref{fig:WS-EP-P-MSD-vs-time}(a,b).
}\label{fig:WS-potential-gr}
\end{center}
\end{figure*}

To demonstrate how the method of collineations performs 
in the search for crystalline domains in multicomponent systems, 
we consider a binary system of particles closely related,
from the perspective of the interactions between the particles,
to the system introduced by Wahnstr{\"o}m \cite{wahnstrom1991molecular}.
This original binary system is equimolar. 
It has been extensively studied previously 
\cite{wahnstrom1991molecular,glotzer2000time,lavcevic2003spatially,
coslovich2007understanding,
malins2013longlivedclusters,royall2015strong,
nishio2024liquidus}.

According to the previous investigations \cite{nishio2024liquidus,pedersen2007crystallization,pedersen2010geometry},
in which the eutectic phase diagram has been determined \cite{nishio2024liquidus}, 
the equimolar system is close to one of the 
two eutectic points for the chosen interactions between the particles.
It was also shown that at composition $BA_2$ the system is expected to congruently 
crystallize into the $MgZn_2$ ($BA_2$) crystal structure, 
which is the hexagonal Laves phase ($C14$, $hP12$, $P6_{3}/mmc$). 
This is a rather complex crystal structure in which there are 12 particles 
in the unit cell of the hexagonal lattice \cite{nishio2024liquidus,pedersen2007crystallization,pedersen2010geometry}.
The system that we investigated is essentially the same system that was studied in 
Ref.~\cite{pedersen2007crystallization,pedersen2010geometry}. 

As we already stated above, our goal here is 
a general demonstration of the methods based on considerations of collineations. 
Thus, it is not our intent to discuss
the structural details of the structures that form.
However, we note that the complexity of the $MgZn_2$ crystal structure indicates that in order to elucidate
the relevant structural features it would be necessary to consider as neighbors not only the nearest neighbors
from the perspective of the PDF, 
but also to consider as neighbors particles at further distances 
(See Fig.~\ref{fig:WS-potential-gr}(B)) Ref.~\cite{pedersen2010geometry}.
This point will be demonstrated below.

In the system that we simulated, 
the numbers of the smaller and larger particles were 
$N_{A} = 18'000$ and $N_{B}=9'000$.
Thus, the system that we studied is noticeably larger 
than the system studied in Ref.~\cite{pedersen2007crystallization}.
The particles interacted through the shifted-force 
Lennard-Jones (LJ) potentials introduced according 
to the expression:
$\widetilde{\phi}(r) = \phi(r) - \phi(r_c) - \phi'(r_c)(r-r_c)$ 
\cite{nicolas1979equation,frenkel2023understanding,toxvaerd2011communication}. 
For the cutoff distance we used: $r_c = 2.5\sigma$, 
where $\sigma$ is either $\sigma_{AA}$, $\sigma_{BB}$, or $\sigma_{AB}$.
The used LJ potentials with their parameters are shown in Fig.~\ref{fig:WS-potential-gr}(A).

The parameters of the smaller $A$-particles were used to define the
lengthscale ($\sigma_{AA}$), the energy and temperature scale $\epsilon_{AA}$, 
and the timescale $\tau = \sqrt{m_{AA}\sigma_{AA}^{2}/\epsilon_{AA}}$.  

We performed NVT simulations using the Nos{\'e}-Hoover 
thermostat within the LAMMPS program.
Simulations were performed in a cubic box with PBC. 
The density of particles was $\rho_o = 0.82\; (1/\sigma_{AA}^3)$, 
as in Ref.~\cite{pedersen2007crystallization}.

In our simulations, the $A$ and $B$ particles initially 
were distributed as alternating planes in the simple cubic lattice. 
This system was quickly heated to $T=3\,\epsilon_{AA}$. 
At this temperature the system immediately melts and 
mixing and equilibration proceed very quickly. 
The mean square displacement of particles at this temperature 
reached more than 1000 $\sigma_{AA}^2$
in our simulations. 
Thus we can safely assume that the system equilibrated at $T=3\,\epsilon_{AA}$. 
This also follows from the monitored dependence of the potential energy on time.
Then the system was brought to $T=1\,\epsilon_{AA}$. 
At this temperature equilibration also happens relatively fast.
On the next step, the temperature was instantaneously reduced
to the actual temperatures of interest. 
Then the evolution of the system was monitored.
We observed crystallization of the system at temperatures 
in the interval $T \in [0.57, 0.62]\; (\epsilon_{AA})$. 
Here we partially describe the results obtained at $T=0.58\epsilon_{AA}$ 
as our goal is to demonstrate the method of collineations 
for the studies of crystallization.

\subsection{Results from the binary system}

\begin{figure*}
\begin{center}
\includegraphics[angle=0,width=6.5in]{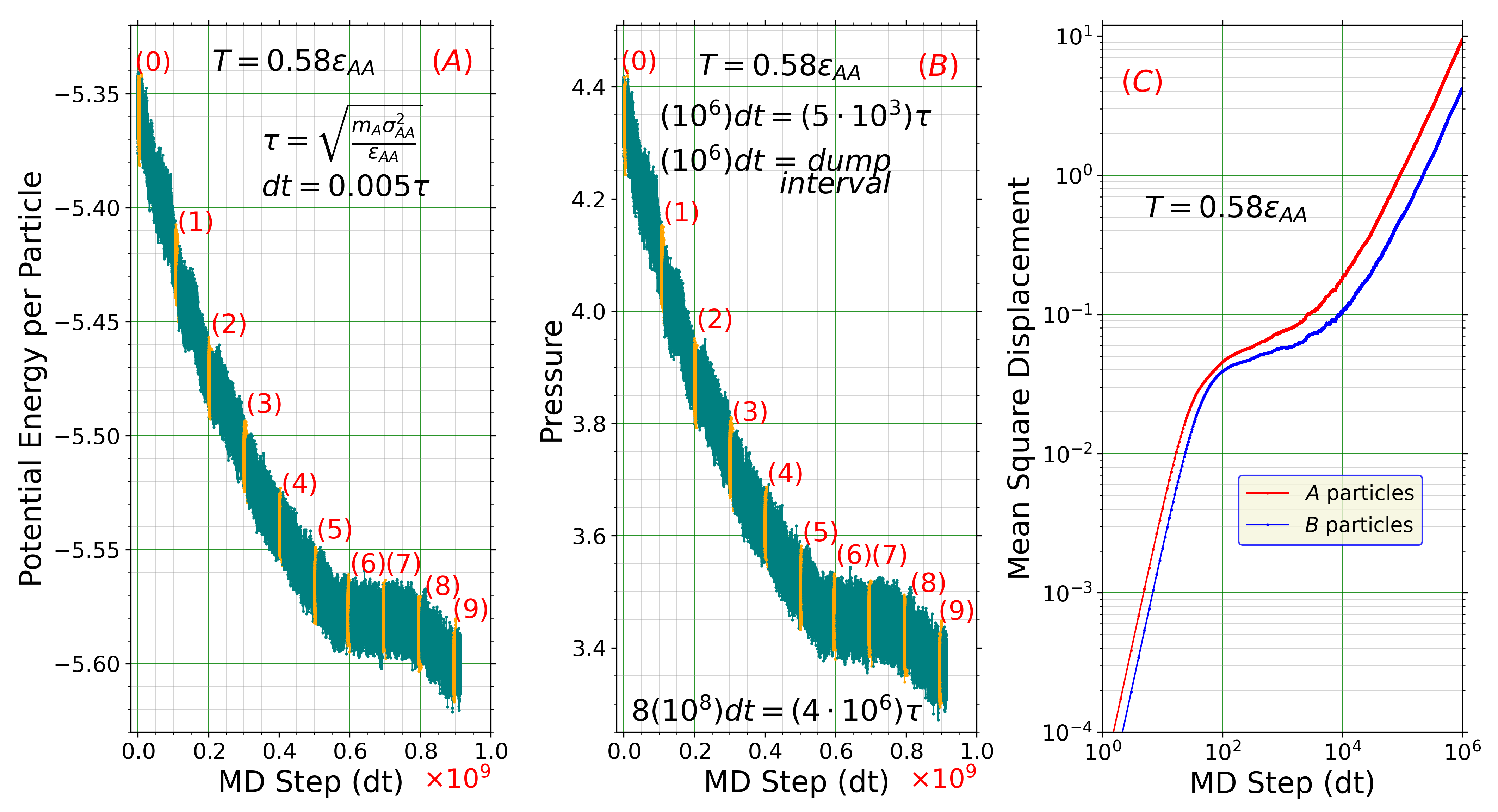}
\caption{
Fig. 8. Panels (A) and (B) show the changes in the system’s 
potential energy and pressure during the crystallization process.
The beginnings of the orange regions mark the times at which configurations shown in Fig.~\ref{fig:Frames-LLL-01}
produced. 
Also, from the orange regions marked as (0), (4), and (9),
containing 10 configurations each, the PPDFs shown in Fig.~\ref{fig:WS-potential-gr} were produced.
Panel (C) displays the dependence of the mean square displacements (MSDs) of particles on time at the starting time corresponding to the beginning 
of the time interval (0) in panels $(A)$ and $(B)$. 
}\label{fig:WS-EP-P-MSD-vs-time}
\end{center}
\end{figure*}

\begin{figure*}
\begin{center}
\includegraphics[angle=0,width=6.5in]{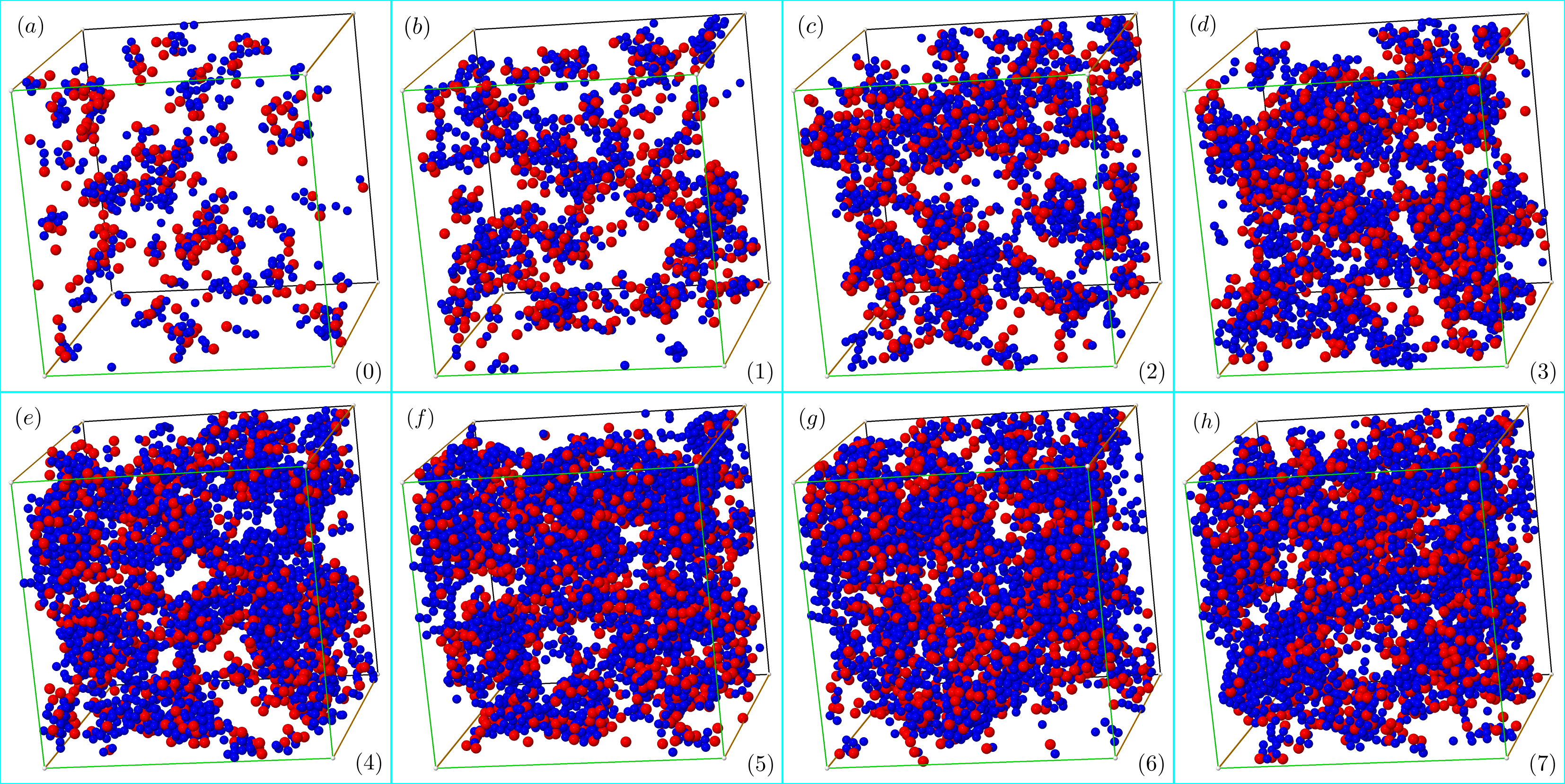}
\caption{Results from the C-PNBR method (\ref{subsec:C-PNBR}). 
Snapshots of the processed configuration 
frames at selected times during the crystallization process.
In the snapshots are shown only those particles that belong to the $C(6,15^{\circ})$ collineations 
and simultaneously have at least 6 neighbors 
that also belong to the same type of collineations. 
The $A-A$, $A-B$, and $B-B$ neighbors were defined according 
to the cutoff distances corresponding to the first minima in the corresponding
PPDFs:
$r_{c,AA} =1.51$, $r_{c,AB} =1.62$, and $r_{c,BB} =1.71$.
The marks from (0) to (7) in the bottom right corners of the snapshots correspond to the moments in time, 
shown in Fig.~\ref{fig:WS-EP-P-MSD-vs-time}(a,b), at which these configurations were produced.
The collineations of blue particles 
are visible to the eye closer to the top-left-front corners of the snapshots
(e,f,g,h). Overall, it is evident from the snapshots that the number of particles that are selected increases as crystallization
progresses. It is also quite clear that the crystallization does not develop through the usual nucleation mechanism.
}\label{fig:Frames-LLL-01}
\end{center}
\end{figure*}

In Fig.\,\ref{fig:WS-EP-P-MSD-vs-time}(a,b) we show the dependencies 
on time of the potential energy and pressure of the system in the process of crystallization. 
The presented data show the results after a short period of initial relaxation
($\Delta t \approx 10^{6}$ MD steps). 
Note that the crystallization proceeds extremely slowly, 
in agreement with the results reported in 
Ref.~\cite{pedersen2007crystallization}.
The slowness of the crystallization process becomes apparent if the results in panel (c)
are considered. Here we clearly observe the caging of the particles, and we also see that on the
timescale of $10^{6}$ MD steps the mean square displacement of the particles reaches $10 \sigma^2$ 
for the small particles and $4 \sigma^2$ for the large particles.

We analyzed the structural data using the collineations of particles.
The $A-A$, $A-B$, and $B-B$ neighbors were defined according 
to the cutoff distances corresponding to the first minima in the corresponding
PPDFs:
$r_{c,AA} =1.51$, $r_{c,AB} =1.62$, and $r_{c,BB} =1.71$. 
The value of the cutoff angle for the chain determination was
$\xi_{max}=15^{\circ}$ for all triples of particles.

In Fig.~\ref{fig:Frames-LLL-01}, we show the snapshots of the processed
configurations at the beginnings of the selected time intervals. 
Only those particles are shown that belong
to the $C(6,15^{\circ})$ collineations and simultaneously have at least 6 neighbors
that also belong to collineations of the same type (C-PNBR). It is necessary to note that the requirement to have at least six neighbors leads to the situation when the presence of the chains in the figures is not apparent, as some particles that belong to the collineations are not displayed.

In Fig.~\ref{fig:Frames-LLL-01} we see that,
shortly after the initial relaxation process,
in the system start to form domains that are
revealed by the method of collineations.
Thus, it is reasonable to assume that in 
these domains some translational order develops that 
has some resemblance to the crystal order.
Note that in the snapshots there are
particles of two types.
It is reasonable to expect that 
in the crystallization mechanism an important role should
play the process of chemical ordering. 
Note also that in Fig.~\ref{fig:Frames-LLL-01} 
we do not observe the nucleation process from the perspective of seeing
one or several somewhat spherical nuclei. 

\begin{figure*}
\begin{center}
\includegraphics[angle=0,width=7.0in]{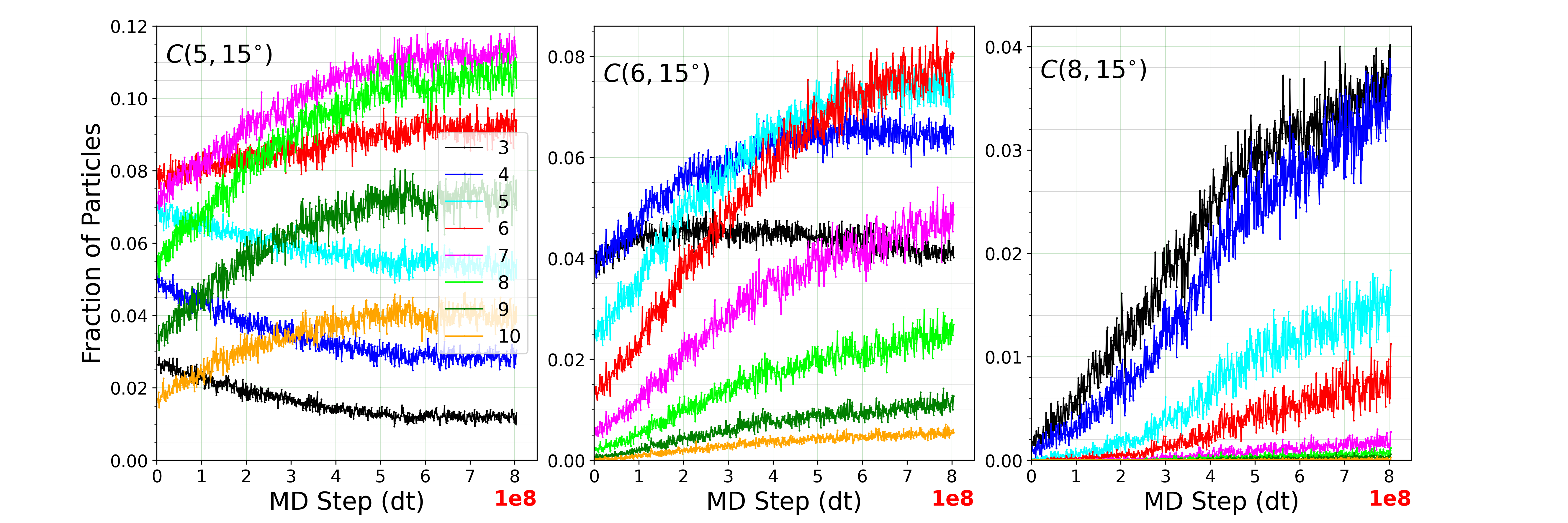}
\caption{The increase in the numbers of particles that belong to collineations of some type and simultaneously
have at least some numbers of neighbors that also belong to collineations of the same type. 
Different panels show the data obtained using the collineations 
of types $C(5,15^{\circ})$, $C(6,15^{\circ})$, and $C(7,15^{\circ})$, correspondingly.
In panel $C(5,15^{\circ})$, the numbers that mark the curves correspond 
to the minimal numbers of collineation-neighbors that are necessary for the particles to be counted.
The curve corresponding to the $C(6,15^{\circ})$ collineations with at least 6 collineation-neighbors 
exhibits a particularly well-pronounced growth. 
Thus, this ``type" of particles was used to produce 
the snapshots in Fig.~\ref{fig:Frames-LLL-01}.
}\label{fig:lll-nbr-increase-with-time}
\end{center}
\end{figure*}

\begin{figure*}
\begin{center}
\includegraphics[angle=0,width=7.0in]{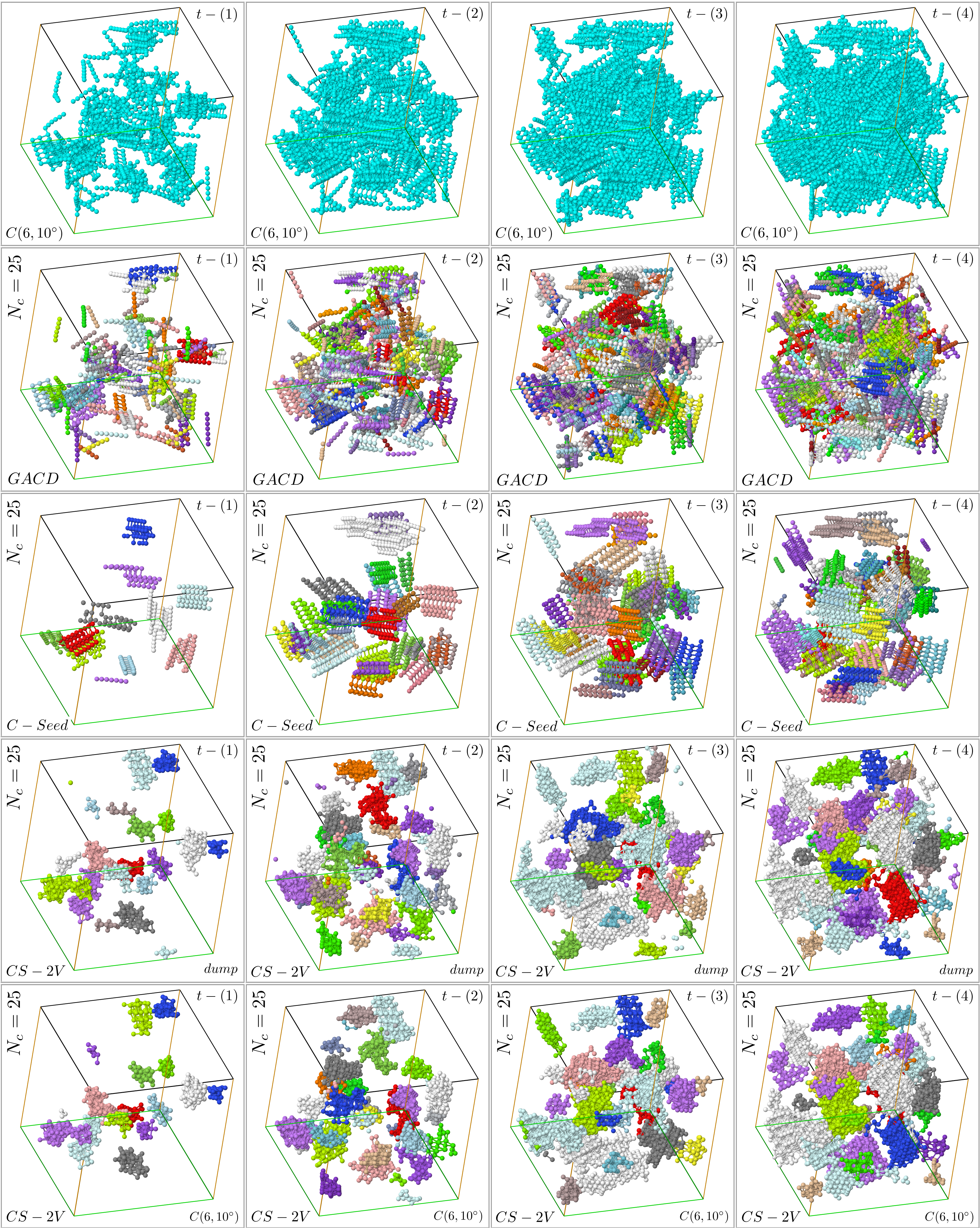}
\caption{
Results for the collineations and crystallites extraction from differet methods
for the smaller $B$ ($Zn$) particles at times (1), (2), (3), (4). See main text for detailed description. Only clusters/crystallites containing at least 25 particlesare shown.
}\label{fig:cryst-extraction-1234}
\end{center}
\end{figure*}

\begin{figure*}
\begin{center}
\includegraphics[angle=0,width=7.0in]{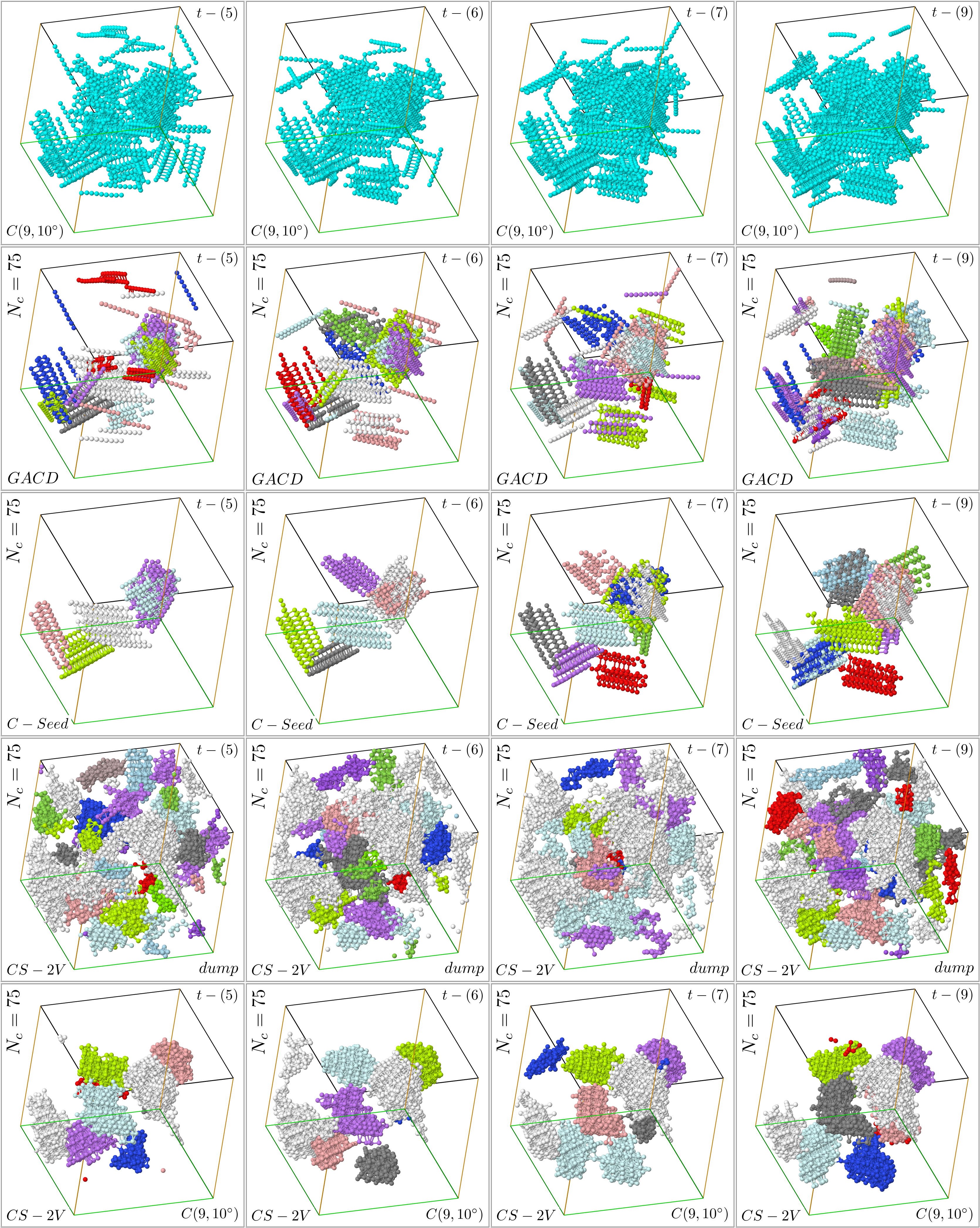}
\caption{
Results for the collineations and crystallites extraction from differet methods
for the smaller $B$ ($Zn$) particles at times (5), (6), (7), (9). See main text for detailed description. Only clusters/crystallites containing at least 75 particlesare shown.
}\label{fig:cryst-extraction-5679}
\end{center}
\end{figure*}

In order to quantify the progress of 
the crystallization process, 
we show  
in Fig.~\ref{fig:lll-nbr-increase-with-time} 
the dependencies on time of the total numbers of particles
that belong to the collineations of particular types and simultaneously have some number of neighbors that also belong to the collineations of similar type. 
It is clear that the shown curves provide some measure
of the crystal organization from the structural perspective.
In the middle panel, corresponding to the 
$C(6,15^{\circ})$ collineations, 
the increase in the number of particles that 
have 6 neighbors that also belong to the $C(6,15^{\circ})$ 
collineations is particularly apparent. 
Thus, these particles were used as the lower bound for the particles 
shown in Fig.~\ref{fig:Frames-LLL-01}. 
If the particles having five neighbors were also included then the
panels in Fig.~\ref{fig:Frames-LLL-01} would be overpopulated, i.e.,
less clear for visual inspection.

Clearly, this approach represents a direct adaptation 
of the method successful for the study of the single-component system
to the binary system with much more complex structural organization.
We note that while the cutoff distance used for the $B-B$ pairs may seem reasonable
from the perspective of PPDFs, it is, in fact, 
not sufficient to observe collineations that 
should be formed by $B$-particles only.
That is because from Ref.~\cite{pedersen2007crystallization} 
we know that we expect to observe a rather complex crystal structure and that
for revealing this structure with the method of collineations it might be necessary to consider
the second neighbors in the PDF.

It is clear that because of translational periodicity, 
collineations must be formed by particles 
that occupy the same basis sites in
different unit cells.
Thus, to observe collineations of the $Mg$ particles in the perfect
$MgZn_2$ type crystal lattice, 
it might be necessary to consider such distances as  $r \sim 2.3 \sigma$.

Also, it is clear that to reveal crystal ordering 
it might be reasonable to consider separately the 
collineations formed by the $A$ and $B$ ($Zn$ and $Mg$) particles. 
That is the approach that we adopt further.

In general, the C-PNBR method is sufficient for monitoring the overall progress of crystallization and also for elucidating 
those particles in the crystallites that can be used for the 
determination of the crystal structure. 
However, if it is of interest to learn about 
the size distributions of the crystallites then this method is not particularly suitable. 
For this purpose, other methods that we secribe in the next subsection are more suitable.

\subsection{Application of the 
GACD (\ref{method:GACD}), 
C-Seed (\ref{subsec:C-Seed}), and 
CS-3V (\ref{subsec:CS-3V}) 
methods to 
the smaller 
$A$-particles \label{subsec:Zn-results}}

In Fig.~\ref{fig:cryst-extraction-1234}
the results for 
the extraction of collineations and crystallites by the
GACD (\ref{method:GACD}), 
C-Seed (\ref{subsec:C-Seed}), and 
CS-3V (\ref{subsec:CS-3V}) 
methods for the smaller $A$ ($Zn$) particles 
are shown. 
In production of the data,
all cutoff angles for all methods on all stages of the data
processing were $10^{\circ}$.

In the 1st row of Fig.~\ref{fig:cryst-extraction-1234} 
only those particles are shown that belong to 
the $C(6,10^{\circ})$ collineations. 
The snapshots in the 1st, 2nd, 3rd, and 4th columns 
correspond to the times (1), (2), (3), (4) 
in Fig.~\ref{fig:WS-EP-P-MSD-vs-time}(a,b).
Note that the development of crystalline ordering is apparent.
Also note that particles extracted in this way already 
can be used for the structural analysis of the 
crystallites that form. 
It is clear that there are several crystallites 
with different orientations.

In the 2nd row of 
Fig.~\ref{fig:cryst-extraction-1234} 
the results of the grouping of collineations 
from the first row into the clusters by the 
$GACD$ (\ref{method:GACD}) method are shown. 
Only clusters containing more than 25 particles are shown. 
For the coloring of the particles the following convention is adopted. 
The largest cluster is associated with the hydrogen atoms, 
the second largest cluster is associated with the helium atoms, 
and so on along the periodic table of elements.
Then, in the visualization program, a particular color
can be assigned to each element. 
In this way different crystallites are colored differently and
different colors do not mean that there are different particles.

It is clear that the $GACD$ method provides certain
separation of crystallites. 
However, it suffers from the fact that it can classify
particles that apparently belong to the same crystallite into two different crystallites. 
This happens because in the current application of the method
the program searches for collineations (directed in all directions) 
starting on particles that are not assigned to any crystallite. 
Note that on this stage each existing crystallite 
is characterized by only one direction---the direction of the founding collineation.
When a new collineation is found it is assigned to an existing crystallite 
or it starts a new crystallite. 
If instead the program would search for all collineations that suit a 
selected crystallite then this effect is likely to be weakened.
However, we do not consider such search as very effective from 
the perspective of the required calculations. 
Despite this shortcoming of the method 
we find this method useful for the structural analysis as an additional reference to which results
from the other methods can be compared.

In the 3rd row 
of Fig.~\ref{fig:cryst-extraction-1234},
the clusters formed by the C-Seed (\ref{subsec:C-Seed}) 
method are shown. 
The C-Seed method was applied to the collineations shown in the 1st row of the corresponding columns.
The collineations used as seeds for a crystallite had to contain at least 5 particles.
Also, only those particles could join the crystallites that themselves belong to the $C(5,10^{\circ})$ collineations. 
Only clusters containing 
more than 25 particles are shown.

From the comparison of the 2nd and 3rd rows of the figure it is clear that 
the C-Seed (\ref{subsec:C-Seed}) method provides a much clearer separation between the crystallites than the $GACD$ (\ref{method:GACD}) method. 
The C-Seed method also clearly exposes the crystalline order within the clusters.
It also can be noted that sometimes the C-Seed method also produces interpenetrating
crystallites which, likely, 
should be classified as one crystallite.
However, 
the C-Seed method clearly finds a smaller number of particles than the GACD method.
It is expectable because the C-Seed method cannot add to a crystallite 
a collineation which is far away from the core of the crystallite, while the GACD method 
can add such a collineation. In our view, it might still be reasonable to compare the results
from the two methods to gain an insight into what one or another method can miss. 
On the other hand, this insight can also be gained by comparing the results from the 
C-Seed method with the first row of the figure, i.e., 
with the unsorted collineations of a particular type.

In the 4th row, 
the results from the CS-2V (\ref{subsec:CS-3V}) crystallites construction 
algorithm applied directly to the structure files (dump files) produced by LAMMPS are shown. 
The number of the directional vectors used (required) for joining two particles 
into a crystallite was two and correspondingly we call the method CS-2V.
In Fig.~\ref{fig:cryst-extraction-1234}, 
only clusters containing at least 25 particles are shown.
If instead of two vectors were considered 3 vectors then the number
of the found particles would be too small to make a reasonable comparison with the 
C-Seed method (for example). With the choice of 2 vectors the comparison 
clearly can be made.
Two observations should be noted. First, the domains found by the CS-2V do not really look as crystalline even though their location and sizes are somewhat similar to the results found by the 
C-Seed method. 
Thus, the results from 
the CS-2V method can be used to estimate the progress of the
crystallization process. This progress is evident from the comparison of 
the proper panels in Fig.~\ref{fig:cryst-extraction-1234},\ref{fig:cryst-extraction-5679}.
However, the actual meaning of the found regions remains obscure. It is also clear from the comparison 
of the panels corresponding to times (3) and (4) for the C-Seed and CS-2V methods that the CS-2V
method finds larger numbers of particles. This observation suggests that the CS-2V method, besides finding the particles in crystallites, also finds particles which are in a pre-crystallite state. 
If this is so then the CS-2V method might allow the investigation of the ordering in the pre-crystallite regions. This point clearly requires further investigation.

It is possible to use the structural output from one of the considered methods as a structural 
input for another method. 
This operation might help to understand relations between different methods. 
We already used this approach because the structural output from the search for collineations 
serves as the structural input for the GACD and C-Seed methods.
It is of interest to consider how the CS-2V method will perform on the output from the collineation search program.
To apply this operation it is necessary to realize that in order to determine the directions associated with each particle that belongs to the collineations it is necessary to consider not only particles that form collineations but also their neighbors.

It is worth noting that if the centro-symmetry crystallites construction algorithm is applied to the collineations of length 5 with their neighbors
then there might be a high probability that the results would be very
similar to the results shown in the 4th row. 
It is likely to be like that because if two neighboring particles have four directional vectors 
which are similar enough (two vectors from each particle) then it is possible that two vectors (one from each neighboring particle) form a single collineation of length 5. We verified this assumption and found that indeed if the CS-2V approach is applied to the collineations of length 5 with their neighbors then the results are essentially indistinguishable from those presented in the 4th row.

In the 5th row, the results from the $CS-2V$ algorithm applied to the particles forming collineations of length 6 (and their neighbors) are shown. It is clear that there are differences with the 4th row. The results in the 5th row can also be compared with the results obtained 
with the C-Seed method.

Fig.~\ref{fig:cryst-extraction-5679} is similar to the 
In the 1st row of Fig.~\ref{fig:cryst-extraction-5679} 
only those particles are shown that belong to the $C(9,10^{\circ})$ collineations. 
The snapshots in the 1st, 2nd, 3rd, and 4th columns correspond to the times (5), (6), (7), (9) in Fig.~\ref{fig:WS-EP-P-MSD-vs-time}(a,b).
,\ref{fig:cryst-extraction-5679} 
In Fig.~\ref{fig:cryst-extraction-5679} 
only clusters containing more than 75 particles are shown.

Fig.~\ref{fig:cryst-extraction-5679} is similar
to Fig.~\ref{fig:cryst-extraction-1234}.
The differences are the following.
In Fig.~\ref{fig:cryst-extraction-5679}
the snapshots in the 1st, 2nd, 3rd, and 4th columns correspond to the times (5), (6), (7), (9) in Fig.~\ref{fig:WS-EP-P-MSD-vs-time}(a,b).
In the first row only those particles 
are shown that belong to the $C(9,10^{\circ})$ collineations. 
Correspondingly, the results of the analyses 
in the second, third, and fifth rows are based
on the input structures formed by the $C(9,10^{\circ})$ collineations.
However, the processing itself for 
the GACD and C-Seed methods, similarly to 
Fig.~\ref{fig:cryst-extraction-1234}, is based on consideration of the $C(5,10^{\circ})$ collineations.
Finally, in Fig.~\ref{fig:cryst-extraction-5679} 
only clusters containing more than 75 particles are shown.

In general, the comparison of all shown results suggests that the C-Seed method clearly demonstrates crystalline order and it also provides a rather good separation of 
different crystallites. However, we find the comparisons with other methods useful. 
Thus, for example, comparison with the
CS-2V method suggests the existence of the pre-crystalline regions.

\section{Conclusion \label{sec:conclusion}}

Here we suggested and discussed several approaches that can be useful, 
in our view, 
in the studies of crystallization and nucleation processes. 
In particular, these methods can be useful when the forming
crystalline ordering is different 
from the FCC, BCC, or HCP orderings.
Especially, these approaches can be suitable for the studies
of multicomponent systems.
All suggested methods are based on considerations 
of collineations of particles.
The suggested methods allow the extraction of the crystalline regions from the liquid environment and (with some limitations) separation of different crystallites for further structural analysis.

To demonstrate the suggested methods we considered 
a single-component system and a binary system. 
The single-component system was used to demonstrate 
how the results from the suggested methods compare 
with the well-established considerations based on bond-orientational order parameters. 
It was shown that the suggested methods perform well.

Then we considered a binary system with particles interacting
through the Lennard-Jones potentials with the Wahnstr\"{o}m values of the parameters. The crystallization process in this system is very slow and apparently rather complex. 
For this case, in our view, the usefulness of the suggested methods was also clearly demonstrated.

The demonstration of the usefulness of the suggested methods was our primary goal. 
However, for the studied systems we made several observations which deserve 
to be mentioned in the conclusion. 
Thus, for the single-component system we studied crystallization in essentially a glass state. 
For this system we noticed that the observed nucleos existed 
in the system for a very long time essentially without growing until it reached some 
size at which it started to grow very fast.
For the studied binary system in a rather viscous state, 
we observed formation of multiple nuclei that after 
the formation exhibited very slow growth.

\section{Supplementary Materials}

The supplementary materials (SM) for this paper contain more details and results
concerning the studied systems. 
In the SM, there are also several programs used in this study with detailed user manuals (written by the LLM DeepSeek).
These programs and manuals can be used, in particular, to develop further programs for related data analysis 
with the assistance of an LLM.

\section{Data Availability Statement}

Some data and programs that support the findings of this study are available within the article and its supplementary materials.
More data that support the findings of this study are available from the corresponding author upon reasonable request.

\section{Acknowledgements} 

We gratefully acknowledge access to the 
Supercomputing Center of Novosibirsk State University (http://nusc.nsu.ru).


\bibliographystyle{zbib-unsrtnat}

\bibliography{Nucleation-Collineations}

\end{document}